\documentclass[prl,print,superscriptaddress,twocolumn]{revtex4-2}
\usepackage{amsmath}
\usepackage{amsfonts}
\usepackage{amssymb}
\usepackage{graphicx}
\usepackage[colorlinks,linkcolor=blue,anchorcolor=blue,citecolor=blue,urlcolor=blue]{hyperref}
\usepackage{mathrsfs}
\usepackage{epsfig}
\usepackage{float}
\usepackage{subfigure}
\usepackage{mathtools}

\begin{document}
\title{Observation of Exact Mobility-Edge Phases in Two-Dimensional Photonic Quasicrystals}

\author{Chaohua Wu}
\affiliation{Quantum Information Institute, School of Physics and Laboratory of Zhongyuan Light,
Zhengzhou University, Zhengzhou 450001, China}

\author{Yingying Shi} 
\affiliation{Quantum Information Institute, School of Physics and Laboratory of Zhongyuan Light, Zhengzhou University, Zhengzhou 450001, China}

\author{Juan Kang}
\email{jkang@zzu.edu.cn}
\affiliation{Quantum Information Institute, School of Physics and Laboratory of Zhongyuan Light,
Zhengzhou University, Zhengzhou 450001, China}

\author{Qinglong Zhang}
\affiliation{Quantum Information Institute, School of Physics and Laboratory of Zhongyuan Light,
Zhengzhou University, Zhengzhou 450001, China}

\author{Shuo Liu}
\affiliation{Quantum Information Institute, School of Physics and Laboratory of Zhongyuan Light,
Zhengzhou University, Zhengzhou 450001, China}

\author{Gang Chen}
\email{chengang971@163.com}
\affiliation{Quantum Information Institute, School of Physics and Laboratory of Zhongyuan Light,
Zhengzhou University, Zhengzhou 450001, China}
\affiliation{State Key Laboratory of Quantum Optics Technologies and Devices, Institute of Laser spectroscopy, Shanxi University, Taiyuan 030006, China}

\date{\today}
\begin{abstract}
Mobility edges, separating localized and extended states in energy, are central to localization physics. While exact mobility edges have been established in several one-dimensional quasiperiodic models, their realization in two dimensions remains largely unexplored. Here, we uncover a general route to exact mobility edges in the two-dimensional Aubry-Andr\'{e} quasicrystal by introducing auxiliary subsystems. Through isospectral reduction, we reveal a hidden self-dual structure that yields exact mobility-edge conditions, separating extended, localized, and partially localized states in the spectrum. Tailoring the auxiliary structures controls the effective couplings and resulting mobility edge configurations, while auxiliary resonances further support flat-band states. In a decorated Lieb-type quasicrystal, we experimentally realize these predicted phases in photonic waveguide lattices and probe their energy-dependent localization through site-selective quench dynamics. Our results provide a versatile platform for controlling localization and can be readily extended to other physical platforms and higher-dimensional systems.
\end{abstract}
\maketitle

\textit{Introduction.}---Localization is a fundamental phenomenon in both quantum and classical wave systems. A paradigmatic example is Anderson localization \cite{R1,R2,R3,R4}, where random disorder suppresses diffusion and produces exponentially localized wave functions \cite{R5,R6,R7,R8}. Dimensionality plays a crucial role in Anderson localization by governing the scaling of quantum-interference effects. For noninteracting particles subject to uncorrelated disorder, arbitrarily weak disorder localizes all eigenstates in one and two dimensions \cite{R9,R10,R11,R12,R13}, whereas localized and extended states can coexist in three dimensions \cite{R14,R15,R16,R17}, separated by a mobility edge (ME) \cite{R14}. The presence of a ME renders localization energy dependent and enables energy-selective transport. Nevertheless, MEs in 3D random media are generally difficult to determine analytically, limiting both a clear understanding of their underlying physics and their systematic control.

\begin{figure}[b]
\centering
\includegraphics[width=8.4cm]{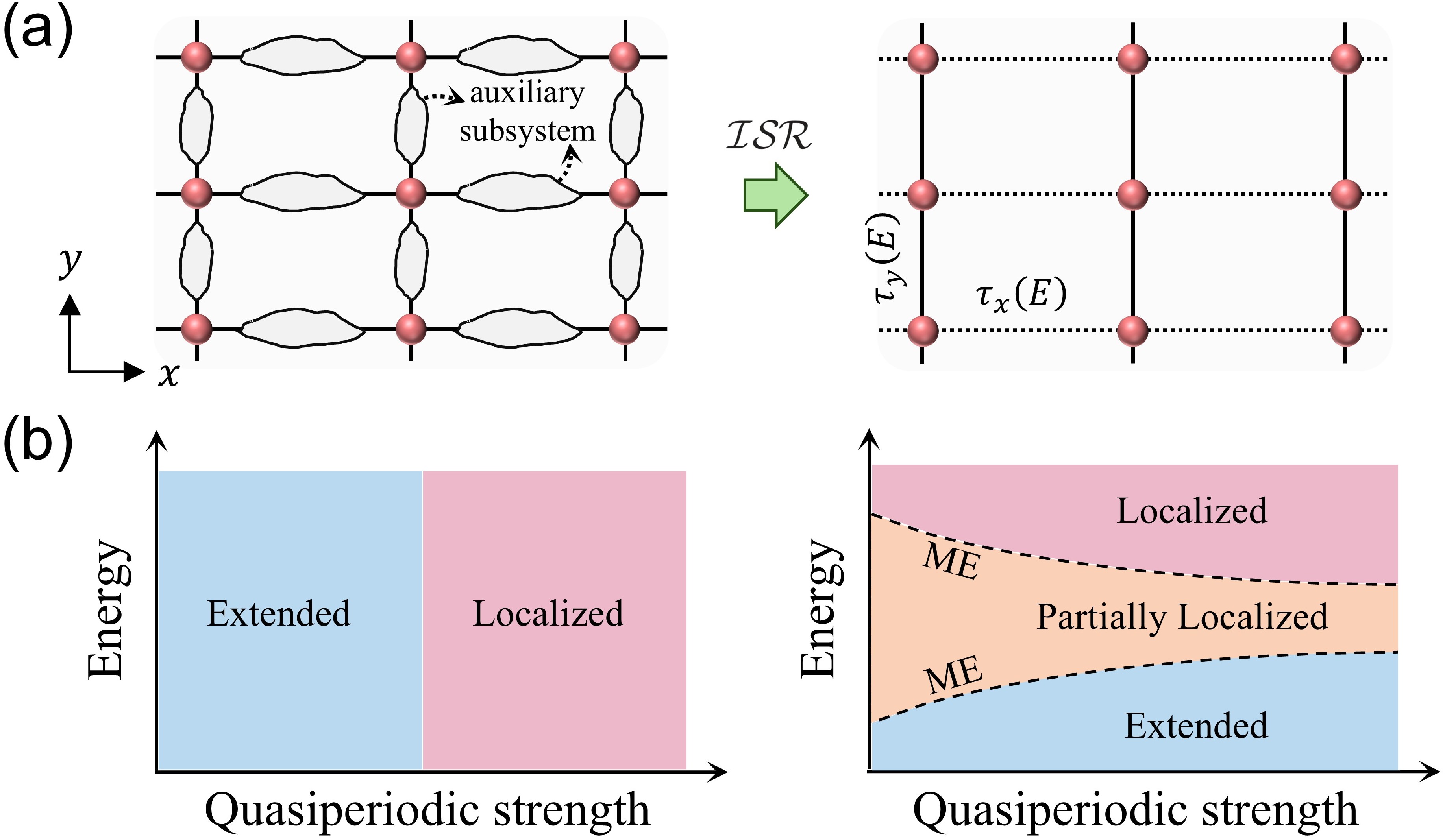}
\caption{(a) Isospectral reduction ($\mathcal{ISR}$) maps the decorated 2D quasicrystal with auxiliary subsystems onto an effective 2D AA model with energy-dependent hoppings $\tau_{x,y}(E)$. (b) Left: in a separable 2D AA model, eigenstates are either extended or localized. Right: the decorated lattice develops MEs that partition the spectrum into extended, partially localized, and localized phases. Partially localized states are localized along one direction but extended along the other.}
\label{F1}
\end{figure}

Quasicrystals \cite{R18,R19,R20,R21}, being neither periodic nor randomly disordered, provide a deterministic platform for exploring rich localization phenomena. The simplest nontrivial example is the Aubry-Andr\'{e} (AA) model \cite{R22}, in which an incommensurate onsite modulation gives rise to a self-dual localization transition shared by all eigenstates \cite{R23,R24}. Generalizations that break this self-dual structure can instead generate MEs even in 1D \cite{R25,R26,R27,R28,R29,R30,R31,R32}, motivating considerable efforts to construct quasiperiodic models with exact MEs \cite{R33,R34,R35,R36,R37,R38,R39,R40}. Most of these developments, however, have been restricted to 1D systems. Increasing dimensionality fundamentally reshapes quasiperiodic localization. In 2D, quasiperiodic scattering becomes intertwined with multidirectional transport, making MEs difficult to establish from simple self-duality conditions. The resulting spectra often exhibit mixed localization behavior \cite{R41,R43,R44,R45,R46}, blurring the boundary between localized and extended states. Although exact MEs have been obtained in specific 2D models \cite{R42}, a general route for constructing 2D quasicrystals with exact MEs remains lacking. Moreover, their experimental observation remains largely unexplored.

In this Letter, we establish a general route to exact MEs in a class of 2D quasicrystals by introducing auxiliary lattice degrees of freedom. Through isospectral reduction, the decorated lattice is mapped onto an effective 2D AA model with energy-dependent couplings [Fig.~\ref{F1}(a)], whose hidden self-duality yields analytically exact ME conditions. By tailoring the auxiliary structures, different ME configurations can be realized, separating extended, localized, and partially localized phases in the spectrum [Fig.~\ref{F1}(b)]. Suitable auxiliary networks can further support resonant flat bands. We realize decorated Lieb-type quasicrystals in femtosecond-laser-written photonic waveguide arrays and, to our knowledge, observe ME phases in two dimensions for the first time. The distinct photonic localization phases are identified through quench dynamics. Our study opens a route to manipulating wave localization in 2D lattices and provides a framework that can be naturally extended to higher-dimensional systems.

\begin{figure*}[t]
\centering
\includegraphics[width=1.8\columnwidth]{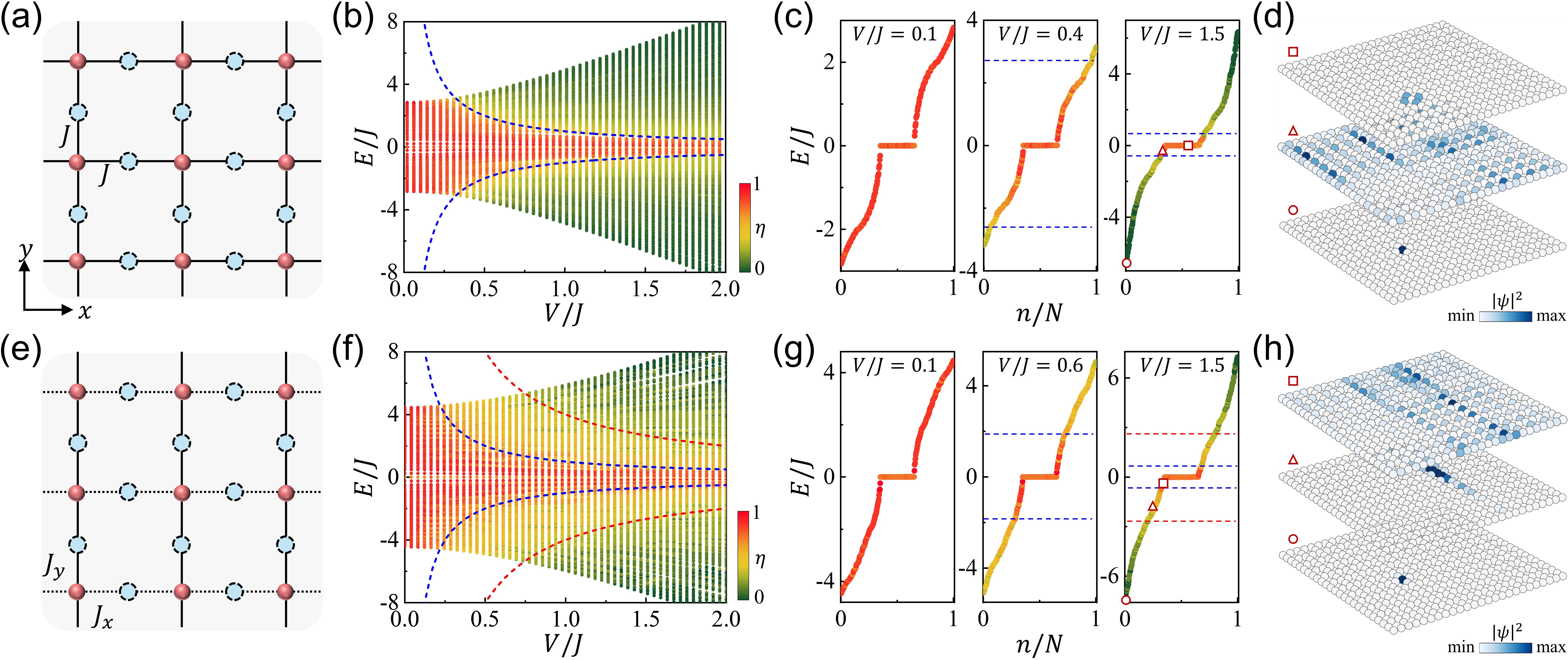}
\caption{(a) Schematic of the decorated Lieb-type quasicrystal with uniform hopping, in which the red and dashed-blue sites denote the quasiperiodic and auxiliary sites, respectively.
(b) Fractal dimension $\eta$ of all eigenstates as a function of energy and quasiperiodic potential strength $V$. The blue dashed lines denote the analytical MEs $E_c=\pm J^2/V$.
(c) Eigenenergies versus the normalized eigenstate index $n/N$ for different values of $V/J$, where $n$ labels the eigenstates and $N$ is the total number of states.
(d) Eigenstate distributions at $V/J=1.5$ for the states marked in (c), corresponding to localized, extended, and flat-band states from bottom to top.
(e)-(h) Similar to (a)-(d), but for anisotropic hopping with $J_x=J_y/2=J$. The states shown in (h) correspond to localized, partially localized, and extended states from bottom to top. The main lattice contains $L_x=L_y=13$ quasiperiodic sites.}
\label{F2}
\end{figure*}

\textit{Hidden self-duality via isospectral reduction.}---
As depicted in Fig.~\ref{F1}(a), we consider a square lattice of quasiperiodic sites 
$\lvert A_{m,n}\rangle$, indexed by the integers $m$ and $n$ along the $x$ and 
$y$ directions, respectively. Neighboring quasiperiodic sites are coupled only 
through periodically repeated auxiliary subsystems. The total Hamiltonian reads
$H=H_{\mathrm{QC}}+H_{\mathrm{aux}}+H_{\mathrm{int}}$,
where $H_{\mathrm{QC}}$ describes the quasiperiodic main lattice, 
$H_{\mathrm{aux}}$ denotes the auxiliary subsystems, and $H_{\mathrm{int}}$ 
represents the couplings between them. The quasiperiodic sector is $H_{\mathrm{QC}}
=\sum_{m,n}V_{m,n}\left\vert A_{m,n}\right\rangle \left\langle A_{m,n}\right\vert $,
with the separable 2D AA potential $V_{m,n}=2[V_x\cos(2\pi\beta_x m)+V_y\cos(2\pi\beta_y n)]$,
where $V_{x,y}$ denote the modulation strengths and $\beta_{x,y}$ are irrational 
parameters characterizing the incommensurate modulations along the two spatial 
directions.

In the basis of the quasiperiodic and auxiliary degrees of freedom, the 
single-particle Schr\"odinger equation takes the block form
\begin{equation}
\left(
\begin{array}{cc}
H_{\text{QC}} & H_{\text{int}} \\
H_{\text{int}}^{\dagger } & H_{\text{aux}}%
\end{array}%
\right) \left(
\begin{array}{c}
\psi _{\text{QC}} \\
\psi _{\text{aux}}%
\end{array}%
\right) =E\left(
\begin{array}{c}
\psi _{\text{QC}} \\
\psi _{\text{aux}}%
\end{array}%
\right) ,
\end{equation}
where $E$ is the eigenenergy of the full system, and $\psi_{\mathrm{QC}}$ and 
$\psi_{\mathrm{aux}}$ denote the wave-function amplitudes on the quasiperiodic 
and auxiliary subspaces, respectively. For 
$E\notin\sigma(H_{\mathrm{aux}})$, with $\sigma(H_{\mathrm{aux}})$ denoting the 
spectrum of the auxiliary Hamiltonian, the auxiliary degrees of freedom can be 
eliminated as $\psi_{\mathrm{aux}}
=(E-H_{\mathrm{aux}})^{-1}H_{\mathrm{int}}^\dagger\psi_{\mathrm{QC}}$,
yielding $[H_{\mathrm{QC}}+H_{\mathrm{int}}(E-H_{\mathrm{aux}})^{-1}H_{\mathrm{int}}^\dagger
]\psi_{\mathrm{QC}}=E\psi_{\mathrm{QC}}$.
This procedure constitutes an isospectral reduction~\cite{R47,R48} of the 
full Hamiltonian onto the quasiperiodic subspace, leading to the energy-dependent 
reduced Hamiltonian
\begin{equation}
\mathcal{H}_{\mathrm{QC}}(E)
=
H_{\mathrm{QC}}
+
H_{\mathrm{int}}
(E-H_{\mathrm{aux}})^{-1}
H_{\mathrm{int}}^\dagger.
\label{e2}
\end{equation}
The second term in Eq.~(\ref{e2}) is the energy-dependent self-energy induced by the 
auxiliary subsystems. The resulting eigenvalue equation can be written as $[E-\Sigma(E)]\psi_{m,n}=\tau_x(E)
\left(
\psi_{m+1,n}+\psi_{m-1,n}
\right)+\tau_y(E)
\left(
\psi_{m,n+1}+\psi_{m,n-1}
\right)
+
V_{m,n}\psi_{m,n}$,
where $\Sigma(E)$ denotes the energy-dependent on-site self-energy, while 
$\tau_x(E)$ and $\tau_y(E)$ are the corresponding effective hoppings mediated 
by the auxiliary subsystems. Their explicit forms are determined by the 
structures and parameters of the corresponding auxiliary subsystems \cite{R49}. 
The reduced problem therefore takes the form of a conventional 2D AA 
model \cite{R41} and reveals a hidden self-dual structure absent in the 
original decorated lattice. Owing to the separability along the two spatial 
directions, the $x$- and $y$-direction dualities can be treated independently, 
yielding the self-dual conditions
\begin{equation}
\left\vert \tau _{x}(E_{c}^{x})\right\vert =\left\vert V_{x}\right\vert ,%
\text{ \ }\left\vert \tau _{y}(E_{c}^{y})\right\vert =\left\vert
V_{y}\right\vert ,
\label{e3}
\end{equation}
which define the exact MEs $E_c^x$ and $E_c^y$, respectively. Crucially, the 
auxiliary subsystems render $\tau_x(E)$ and $\tau_y(E)$ energy dependent, 
allowing the self-dual conditions to be satisfied at different energies and 
thereby producing direction-resolved MEs.

Beyond ME engineering, the auxiliary subsystems in 2D quasicrystals can also 
support resonant flat bands \cite{R49}. Such states arise at resonant energies 
$E\in\sigma(H_{\mathrm{aux}})$ for which an auxiliary eigenmode $\lvert\Phi\rangle$ satisfies $H_{\mathrm{aux}}\lvert\Phi\rangle
=
E\lvert\Phi\rangle$ and $H_{\mathrm{int}}\lvert\Phi\rangle
=0$.
Together, these conditions ensure that the resonant auxiliary mode is decoupled 
from the quasiperiodic lattice through destructive interference. For auxiliary 
chains without local dark modes, the flat-band states generally emerge when 
the $x$- and $y$-oriented auxiliary subsystems share a common resonant energy, $\mathcal{D}_x(E)=\mathcal{D}_y(E)=0$, where $\mathcal{D}_{\mu}(E)
=\det\left(E-h_{\mathrm{aux}}^{\mu}\right)$
and $h_{\mathrm{aux}}^{\mu}$ denotes the local auxiliary Hamiltonian along 
direction $\mu$. For multipath cluster decorations, by contrast, destructive 
interference can generate a local dark mode within a single decorated bond, 
yielding a flat band without requiring a common resonance along the two directions \cite{R49}.

\textit{Exact ME phases in decorated Lieb-type 2D quasicrystals.}---
To exemplify the above approach, we consider a simple decoration with one auxiliary site inserted into each nearest-neighbor bond along both spatial directions [Fig.~\ref{F2}(a)]. The resulting geometry resembles an extended Lieb lattice~\cite{R42,R50,R51,R52}, with the quasiperiodic modulation
applied only to the main sites with $V_x=V_y=V$, while the bond-centered auxiliary sites have zero onsite potential. Here, $\beta_x=(\sqrt{5}-1)/2$ and $\beta_y=\sqrt{2}/2$ are chosen for simplicity. We first consider the isotropic case with uniform nearest-neighbor coupling $J_x=J_y=J$. Isospectral reduction then yields identical effective hoppings along the two directions,
$\tau_x(E)=\tau_y(E)=J^2/E$,
so that the two self-dual conditions coincide and give the exact MEs
$E_c=\pm J^2/V$.

To characterize the localization properties, Fig.~\ref{F2}(b) shows the energy spectrum as a function of the quasiperiodic strength $V$, with each eigenstate colored by its fractal dimension
$\eta=-\ln(\mathrm{IPR})/{\ln N}$.
Here, $\mathrm{IPR}=\sum_{i,j}|\psi_{ij}|^4$
is the inverse participation ratio and $N$ denotes the total number of lattice sites. Extended and localized states are characterized by $\eta\to1$ and $\eta\to0$, respectively. As can be seen, the analytical MEs (blue dashed lines) accurately delineate the boundaries separating the extended and localized states. Figure~\ref{F2}(c) further presents the eigenvalue spectra at different values of $V/J$, directly revealing the coexistence of extended and localized states at different energies. In addition, a highly degenerate flat band remains pinned at the resonant energy $E=0$, originating from dark-state combinations of resonant auxiliary modes that destructively decouple from the quasiperiodic main lattice. Representative eigenstate distributions are shown in Fig.~\ref{F2}(d), displaying localized, extended, and flat-band states. These results directly confirm the predicted MEs and the coexistence of resonant flat-band states.

We next turn to the anisotropic case with $J_x=J$ and $J_y=2J$, as shown in Fig.~\ref{F2}(e). The resulting effective couplings become
$\tau_x(E)=J^2/E$, and
$\tau_y(E)=4J^2/E$,
leading to two distinct sets of MEs, $E_c^x=\pm J^2/V$ and
$ E_c^y=\pm4J^2/V$.
These critical energies partition the spectrum into three localization regimes. States with $|E|<|E_c^x|$ are fully extended, whereas those with $|E_c^x|<|E|<|E_c^y|$ are localized along $x$ but remain extended along $y$, forming a partially localized phase. For $|E|>|E_c^y|$, the states become fully localized. These predictions are confirmed by the numerical results in Fig.~\ref{F2}(f), where the analytical MEs (colored dashed lines) accurately delineate the three regimes. Figure~\ref{F2}(g) shows the spectra at $V/J=0.1$, $0.6$, and $1.5$, with the corresponding fractal dimensions $\eta$ distinguishing the three regimes. Representative eigenstates in Fig.~\ref{F2}(h) provide a direct real-space picture of these phases. The resonant flat band at $E=0$ also persists in this case.

More generally, the auxiliary subsystems provide a flexible means of engineering the effective spectral response. Different auxiliary structures and parameters generate distinct energy-dependent effective couplings, leading to diverse exact ME configurations. Their internal resonances further control the emergence and energies of flat bands, enabling simultaneous tuning of localization and dispersion. More examples are provided in the Supplemental Material \cite{R49} and End Matter.

\begin{figure*}[t]
\centering
\includegraphics[width=1.9\columnwidth]{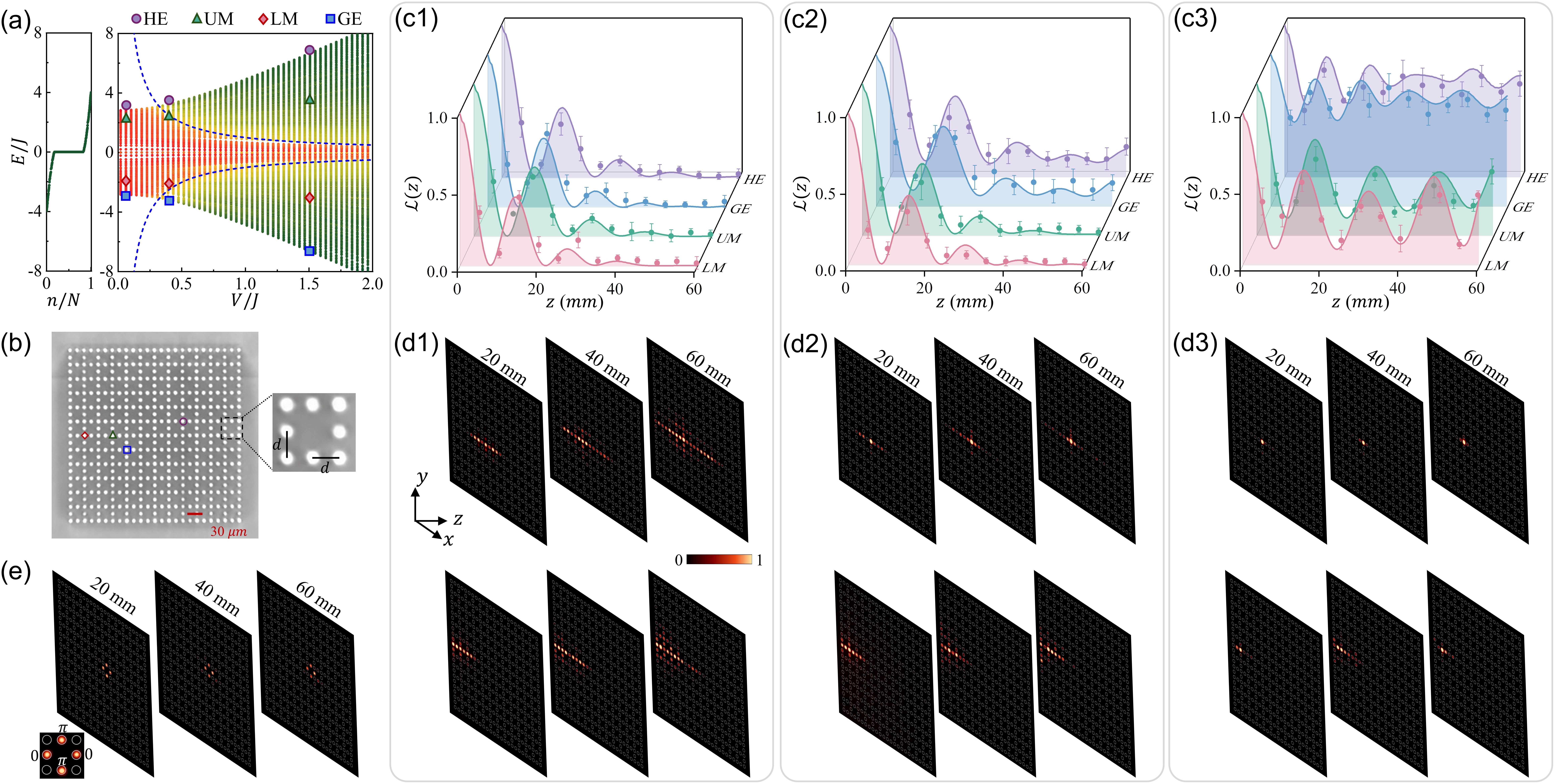}
\caption{Experimental observation of ME phases for $J_x=J_y=J$.
(a) Left: energy spectrum in the decoupled limit $J=0$ at $V=1$. Right: energy spectrum as a function of $V/J$, with the symbols marking the postquench states around highest-energy (HE), upper middle-energy (UM), lower middle-energy (LM), and ground-energy (GE) inputs at $V/J=0.1$, $0.4$, and $1.5$.
(b) Micrograph of the fabricated photonic waveguide array with lattice spacing $d=14~\mu\mathrm{m}$, with the marked waveguides indicating the corresponding input positions.
(c1)-(c3) Measured Loschmidt echoes as a function of propagation distance $z$ for single-site inputs associated with the HE, UM, LM, and GE at $V/J=0.1$, $0.4$, and $1.5$, respectively.
(d1)-(d3) Corresponding output-intensity distributions for the cases in (c1)-(c3), measured at $z=20$, $40$, and $60~\mathrm{mm}$ for the LM and GE excitations. (e) Measured output-intensity distributions for flat-band excitation (lower-left panel) at different propagation distances.}
\label{F3}
\end{figure*}

\textit{Photonic observation of 2D ME phases.}---
To experimentally observe the predicted ME phases, we realize the 2D quasicrystals shown in Fig.~\ref{F2} using femtosecond-laser-written photonic waveguide arrays \cite{R53,R54,R55}. Under the paraxial approximation, light propagation along the longitudinal coordinate $z$ maps directly onto the tight-binding dynamics, with waveguide propagation-constant shifts playing the role of onsite potentials and evanescent couplings realizing the hopping amplitudes. Accordingly, the quasiperiodic onsite modulation is controlled by the laser writing speed, while the coupling strengths are tuned through the inter-waveguide separations. Detailed experimental parameters are provided in the Supplemental Material \cite{R49}.

To experimentally probe the energy-dependent localization, we employ a quench protocol based on site-selective excitation \cite{R40,R56,R57,R58}. Before the quench, the 2D quasicrystal is fully decoupled, such that excitation of a site $\lvert\psi_{i,j}\rangle$ selects an energy determined by its onsite potential. After a sudden quench to the target Hamiltonian $H_f$, the initial state evolves under $H_f$, probing the localization properties of the corresponding energy sector of the postquench spectrum. The localization dynamics are then characterized by the Loschmidt echo, $\mathcal{L}(z)=\left|\langle\psi_{i,j}\rvert e^{-iH_f z}\lvert\psi_{i,j}\rangle\right|^2$, which quantifies the revival probability at the initially excited site. If the postquench energy window lies in an extended region, light spreads away from the initially excited site and $\mathcal{L}(z)$ can approach zero. In contrast, for a localized region, light remains concentrated near the input site, giving a finite $\mathcal{L}(z)$. Repeating the quench for different input sites allows different ME phases to be identified. In our experiment, we trace the single-site quench dynamics at an excitation wavelength of $810~\mathrm{nm}$.

We first experimentally examine the isotropic lattice. In the decoupled limit, four single-site localized eigenstates at distinct energies are chosen as initial states, namely the highest-energy (HE), upper- and lower-middle-energy (UM and LM), and ground-energy (GE) states. The LM and UM states correspond to the 45th and 436th eigenstates, respectively. These states are subsequently quenched to target Hamiltonians with $V/J=0.1$, $0.4$, and $1.5$, with their corresponding postquench spectral positions marked in Fig.~\ref{F3}(a). The fabricated photonic lattice, with a uniform spacing of $d=14~\mu\mathrm{m}$ corresponding to a coupling coefficient of $1~\mathrm{cm}^{-1}$, is shown in Fig.~\ref{F3}(b). Experimentally, the HE, UM, LM, and GE initial states are prepared by injecting light into waveguides $(17,15)$, $(7,13)$, $(3,13)$, and $(9,11)$, respectively. For $V/J=0.1$, the postquench spectral weights of all four inputs lie in the extended region. Accordingly, their measured Loschmidt echoes $\mathcal{L}(z)$ approach zero during propagation [Fig.~\ref{F3}(c1)], as light spreads away from the initially excited sites. At $V/J=0.4$, the MEs cut through the spectrum, with postquench modes around the UM and LM inputs remaining extended, while those around the HE and GE inputs become localized. Correspondingly, $\mathcal{L}(z)$ for the UM and LM excitations approaches zero, whereas that for the HE and GE excitations remains finite [Fig.~\ref{F3}(c2)]. This energy-selective contrast provides a direct dynamical signature of the ME phase. Upon increasing $V/J$ to $1.5$, all four postquench states become localized, with finite $\mathcal{L}(z)$ throughout the propagation [Fig.~\ref{F3}(c3)]. The corresponding output-intensity distributions in Figs.~\ref{F3}(d1)-\ref{F3}(d3) further illustrate these distinct localization behaviors and agree well with the analytical ME predictions. Finally, by simultaneously exciting four auxiliary sites with opposite phases, we prepare a resonant flat-band state whose intensity remains tightly confined throughout propagation [Fig.~\ref{F3}(e)], demonstrating the existence of flat-band localization.

\begin{figure}[t]
\centering
\includegraphics[width=8.4cm]{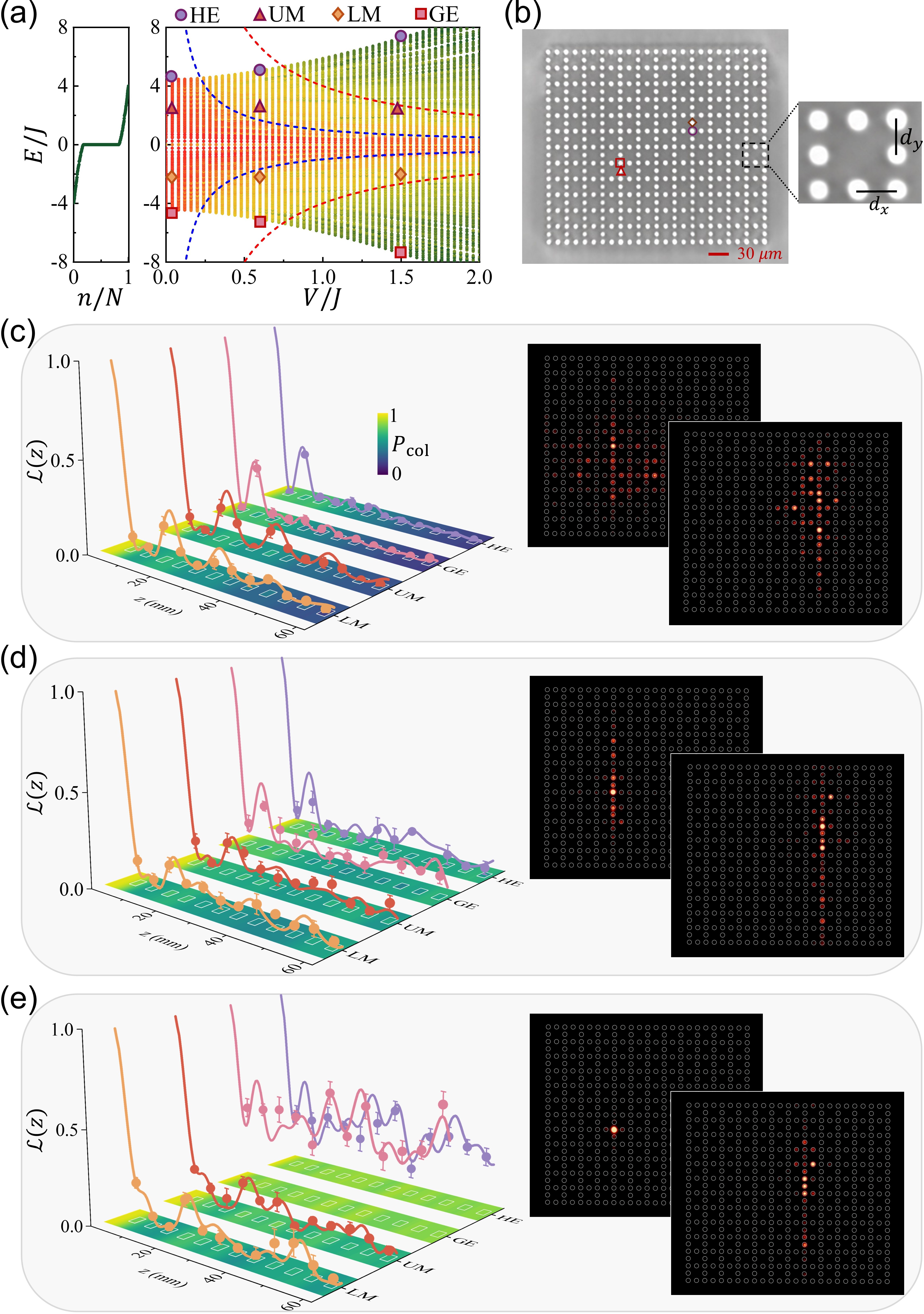}
\caption{Experimental observation of ME phases for $J_x=J_y/2=J$.
(a) Left: energy spectrum in the decoupled limit $J=0$ at $V=1$. Right: energy spectrum as a function of $V/J$, with the symbols marking the selected postquench states around HE, UM, LM, and GE inputs at $V/J=0.1$, $0.6$, and $1.5$.
(b) Micrograph of the fabricated photonic waveguide array with lattice spacings $d_x=14~\mu\mathrm{m}$ and $d_y=12.2~\mu\mathrm{m}$.
(c) Left: measured Loschmidt echoes versus propagation distance $z$ for single-site inputs associated with the HE, UM, LM, and GE spectral regions at $V/J=0.1$. The color map on the base plane represents the column-integrated intensity $P_{\mathrm{col}}$ along the column containing the initially excited site. Right: measured output-intensity distributions at $z=60~\mathrm{mm}$ for the LM (top) and GE (bottom) excitations.
(d), (e) Same as (c), but for $V/J=0.6$ and $1.5$, respectively.}
\label{F4}
\end{figure}

For the anisotropic lattice with the spectrum shown in Fig.~\ref{F4}(a), we probe the quench dynamics using a waveguide array with $d_x=14~\mu\mathrm{m}$ and $d_y=12.2~\mu\mathrm{m}$ (corresponding to a coupling coefficient of $2~\mathrm{cm}^{-1}$) [Fig.~\ref{F4}(b)]. The HE, UM, LM, and GE initial states are prepared by injecting light into waveguides $(17,15)$, $(9,10)$, $(17,16)$, and $(9,11)$, respectively. To characterize the directional localization, we introduce the column-integrated intensity $P_{\mathrm{col}}(z)=\sum_n I_{m_0,n}(z)$, which measures the total optical intensity in the waveguide column containing the initially excited site. At $V/J=0.1$, all four postquench states are extended, as evidenced by the vanishing $\mathcal{L}(z)$ and spreading intensity profiles [Fig.~\ref{F4}(c)]. At $V/J=0.6$, the postquench states enter the partially localized regime. While $\mathcal{L}(z)$ approaches zero as light spreads along $y$, a finite $P_{\mathrm{col}}(z)$ indicates that the excitation remains confined along $x$ [Fig.~\ref{F4}(d)]. At $V/J=1.5$, the postquench states around the UM and LM energies remain partially localized, whereas those around the HE and GE energies become fully localized, exhibiting finite $\mathcal{L}(z)$ and confined intensity profiles [Fig.~\ref{F4}(e)]. These measurements directly demonstrate the direction-dependent MEs.

\textit{Conclusions.}---
In summary, we establish a general framework for exact MEs in 2D quasicrystals through auxiliary-lattice engineering. Isospectral reduction maps the decorated lattice onto an effective 2D AA model with energy-dependent hoppings, whose hidden self-duality gives analytical ME conditions. Tailoring the auxiliary subsystems enables diverse ME configurations, while suitable resonances can additionally generate tunable flat bands. To experimentally demonstrate MEs in 2D quasicrystals, we realize decorated Lieb-type photonic lattices and probe their localization properties using site-selective quench dynamics. Our results provide a versatile framework for engineering localization and flat-band states in 2D quasicrystals, opening new possibilities for their systematic control in more complex and higher-dimensional quasicrystals.

This work is supported by the National Key R \& D Program of China (Grant No. 2022YFA1404500), the National Natural Science Foundation of China (Grant No. 12125406, No. 12404372, No. 12474340), and the Cross-disciplinary Innovative Research Group Project of Henan Province (Grant No. 232300421004).

The data that support the findings of this article are not publicly available. The data are available from the authors upon reasonable request.


\vspace{\baselineskip}

\onecolumngrid

\begin{center}
    \textbf{END MATTER}
\end{center}

\twocolumngrid

Beside the auxiliary subsystems considered in Fig.~\ref{F2}, the isospectral-reduction framework applies to a broad class of auxiliary networks. Their geometry, onsite energies, and internal couplings determine the energy-dependent effective hopping $\tau_\mu(E)$ and on-site energy $\Sigma_\mu(E)$. Representative examples are summarized in Table~\ref{T1}, together with the corresponding $\tau(E)$ and auxiliary spectra $\sigma(h_{\rm aux})$. Because the auxiliary units are periodically repeated, $\Sigma_\mu(E)$ is spatially uniform and can be absorbed into the effective eigenenergy. Consequently, different auxiliary structures generate distinct energy dependences of $\tau_\mu(E)$ and hence different exact ME trajectories through Eq.~(\ref{e3}).

The full auxiliary Hamiltonian is a direct sum of independent bond Hamiltonians,
\begin{equation}
H_{\text{aux}}=\left( \oplus h_{\text{aux}}^{x}\right) \oplus \left( \oplus
h_{\text{aux}}^{y}\right) .
\end{equation}
Accordingly, its resonant energies are determined by the spectra of the local
$x$- and $y$-direction auxiliary units. An auxiliary eigenmode at energy $E$
therefore satisfies $
h_{\mathrm{aux}}^{\mu}|\phi_{\mu}\rangle
=E|\phi_{\mu}\rangle$.
A resonant flat band, however, requires such an auxiliary mode to be
simultaneously decoupled from the quasiperiodic main lattice. For single-path
auxiliary chains, this generally requires a common resonance of the $x$- and
$y$-oriented auxiliary sectors together with destructive interference at the
main sites. Multipath auxiliary networks, by contrast, can support local dark
modes within a single bond unit, allowing flat bands to form without a common
resonance between the two directions.

\begin{figure*}[t]
\centering
\includegraphics[width=1.95\columnwidth]{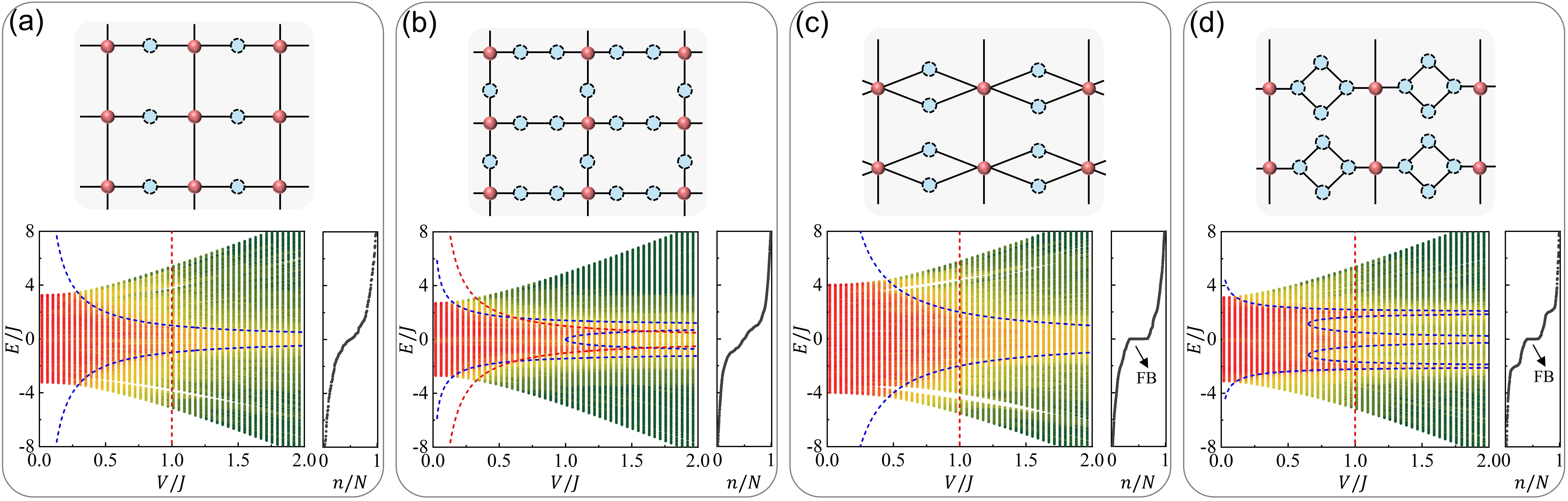}
\caption{Localization properties of representative decorated 2D quasicrystals. The upper panels in (a)–(d) show the corresponding lattice geometries, where the red and blue circles denote the quasiperiodic main sites and auxiliary sites, respectively. The lower panels show the energy spectra as a function of the quasiperiodic potential strength $V/J$, with each eigenstate colored according to its fractal dimension $\eta$. The blue and red dashed curves indicate the analytically predicted mobility edges associated with localization transitions along the $x$ and $y$ directions, respectively. The right subpanels show the eigenenergies at $V/J=2$ as a function of the normalized eigenstate index $n/N$, with the flat bands (FBs) marked where present. The main lattice contains $L_x=L_y=13$ quasiperiodic sites along the two spatial directions. All hopping amplitudes are taken to be uniform with strength $J$.}
\label{F5}
\end{figure*}

Figure~\ref{F5} presents several examples of 2D quasicrystals with uniform hoppings. In Figs.~\ref{F5}(a) and \ref{F5}(b), different single-path auxiliary chains are introduced along the two spatial directions, yielding distinct energy-dependent effective couplings. Specifically, Fig.~\ref{F5}(a) gives $\tau_x(E)=J^2/E$ and $\tau_y(E)=J$, whereas Fig.~\ref{F5}(b) yields $\tau_x(E)=J^3/(E^2-J^2)$ and $\tau_y(E)=J^2/E$, resulting in different exact ME structures. Since the two auxiliary sectors in either model share no common resonance satisfying $\mathcal{D}_x(E)=\mathcal{D}_y(E)=0$, no resonant flat band is formed. In Fig.~\ref{F5}(c), a two-path auxiliary unit supports a local dark mode through destructive interference, giving rise to a flat band at $E=0$ together with $\tau_x(E)=2J^2/E$. For the four-site diamond in Fig.~\ref{F5}(d), $\tau_x(E)=2J^4/[E(E^2-4J^2)]$, with auxiliary resonances at $E=0$ and $E=\pm2J$. Only the zero-energy resonance satisfies the decoupling condition and forms a flat band, whereas the $E=\pm2J$ modes remain hybridized with the main lattice. These examples show that auxiliary-network engineering provides independent control over the exact ME structure and the emergence of resonant flat bands.

\begin{table*}[t]
\centering
\caption{
Effective couplings and eigenenergies of the auxiliary subsystems for different bond decorations.
$J_i$ denote the hopping amplitudes and $U_i$ the onsite potentials of the auxiliary sites.
}
\label{tab:auxiliary}
\renewcommand{\arraystretch}{2.5}
\setlength{\tabcolsep}{8pt}

\resizebox{\textwidth}{!}{
\begin{tabular}{c|c|c|c|c}
\hline\hline
&
\includegraphics[width=0.15\textwidth]{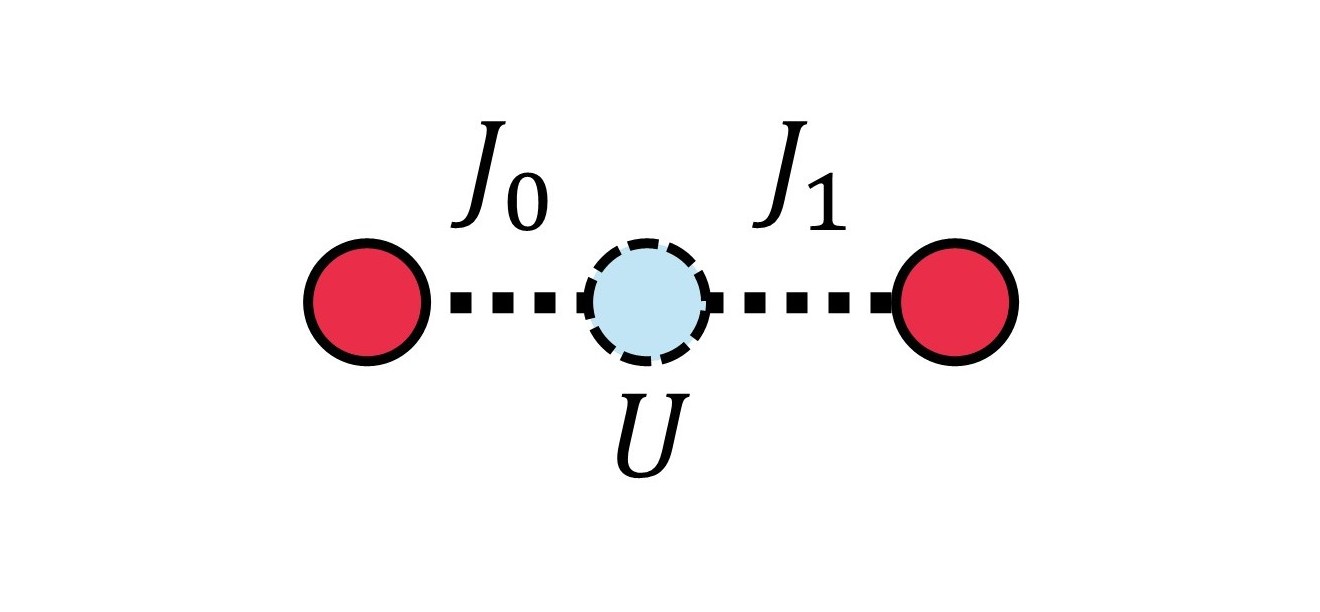}
&
\includegraphics[width=0.15\textwidth]{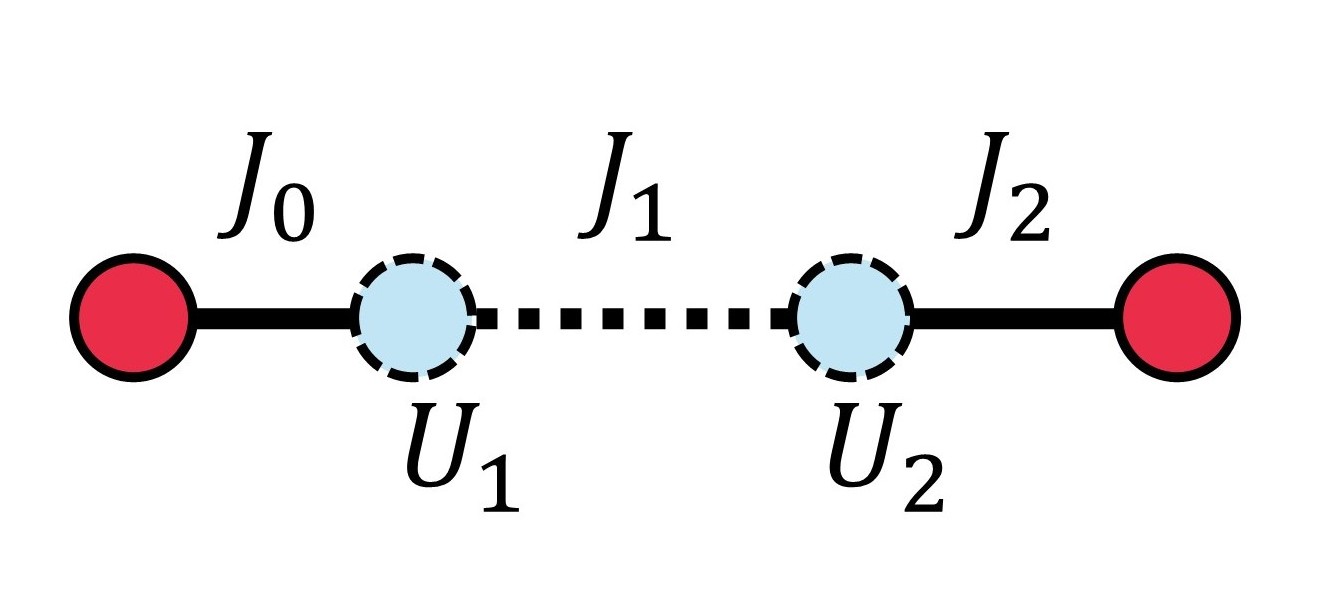}
&
\includegraphics[width=0.15\textwidth]{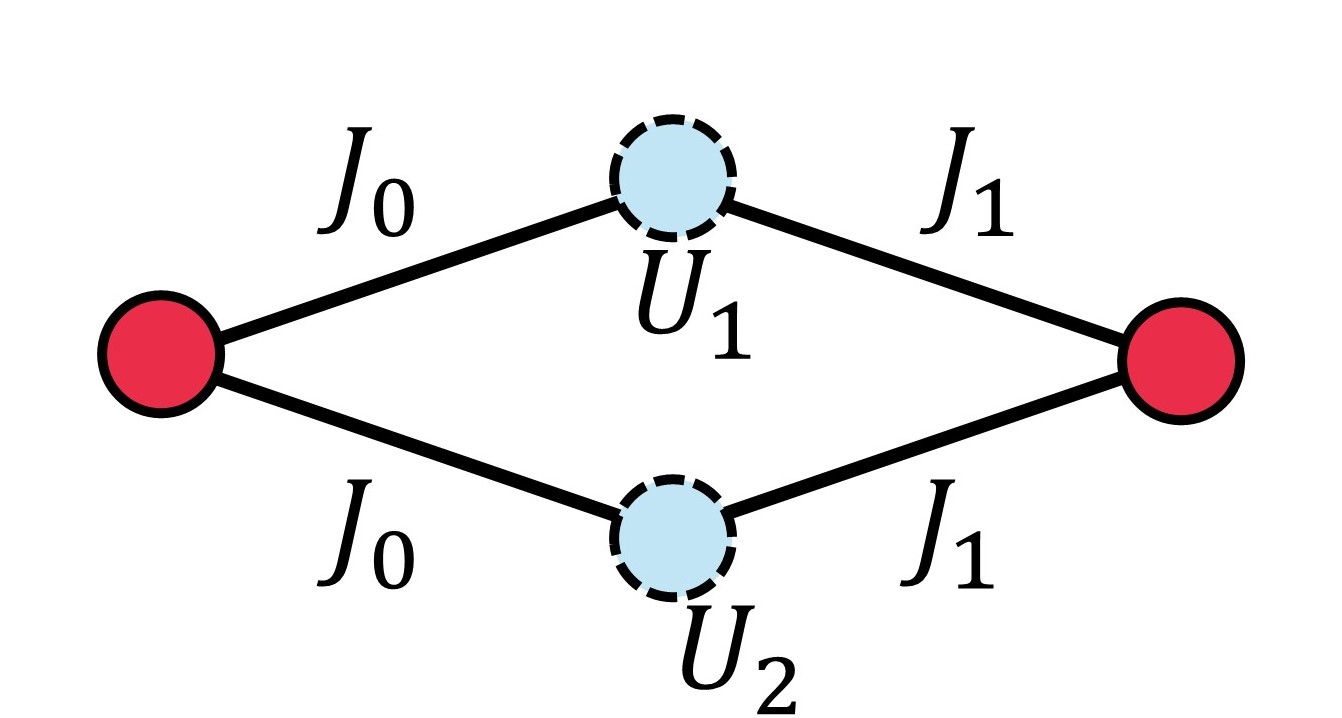}
&
\includegraphics[width=0.15\textwidth]{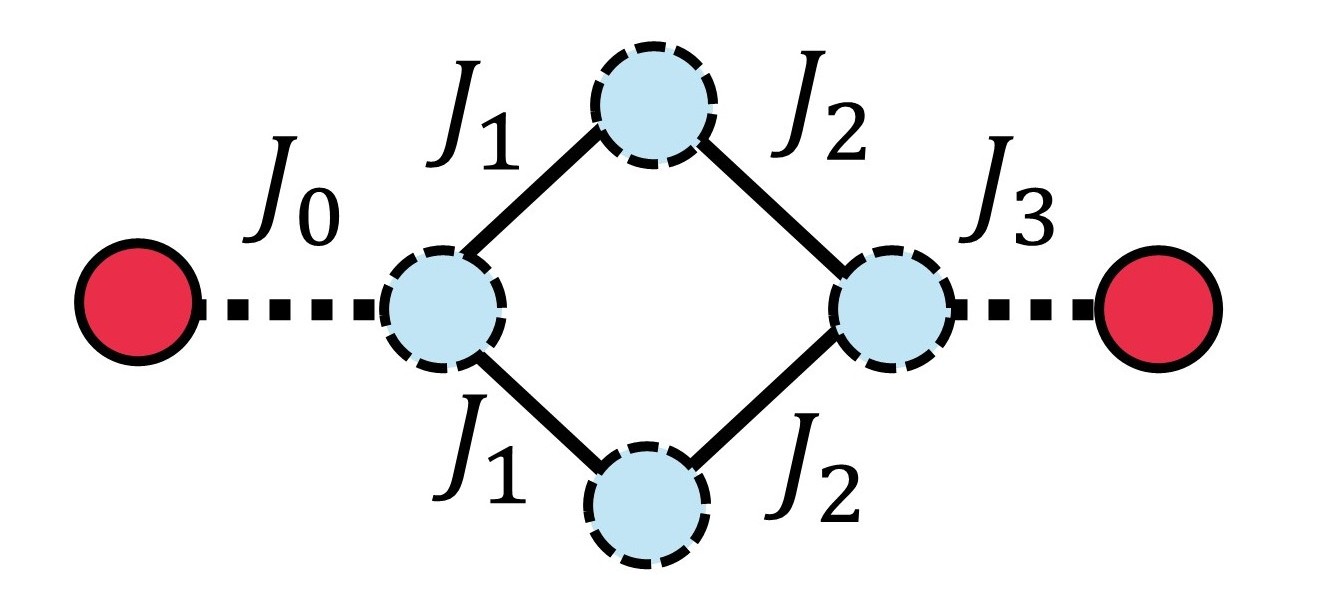}
\\
\hline
\begin{tabular}{c}
$\tau(E)$
\end{tabular}
&
$\displaystyle
\frac{J_0J_1}{E-U}
$
&
$\displaystyle
\frac{J_0J_1J_2}{(E-U_1)(E-U_2)-J_1^2}
$
&
$\displaystyle
J_0J_1
\left(
\frac{1}{E-U_1}
+
\frac{1}{E-U_2}
\right)
$

&
$\displaystyle
\frac{2J_0J_1J_2J_3}
{E\left[E^2-2(J_1^2+J_2^2)\right]}
$
\\
\hline

$\sigma(h_{\mathrm{aux}})$

&
$\displaystyle
\{U\}
$

&
$\displaystyle
\left\{
\frac{U_1+U_2}{2}
\pm
\frac{1}{2}
\sqrt{(U_1-U_2)^2+4J_1^2}
\right\}
$

&
$\displaystyle
\{U_1,U_2\}
$

&
$\displaystyle
\left\{
0,\,
\pm\sqrt{2(J_1^2+J_2^2)}
\right\}
$
\\
\hline\hline
\end{tabular}
}
\label{T1}
\end{table*}

\end{document}